\documentstyle[11pt,aaspp4]{article}

\input psfig

\def\gtorder{\mathrel{\raise.3ex\hbox{$>$}\mkern-14mu
             \lower0.6ex\hbox{$\sim$}}}
\def\ltorder{\mathrel{\raise.3ex\hbox{$<$}\mkern-14mu
             \lower0.6ex\hbox{$\sim$}}}

\begin{document}

\title{Direct Detection of CDM Substructure}

\author{N. Dalal$^1$ and C.S. Kochanek$^2$}
\affil{$^1$Physics Dept., UCSD 0350, 9500 Gilman Dr., La Jolla CA 92093}
\affil{$^2$Harvard-Smithsonian Center for Astrophysics, 60 Garden St.,
  Cambridge, MA 02138}
\affil{email: ndalal@ucsd.edu, ckochanek@cfa.harvard.edu}

\begin{abstract}
We devise a method to measure the abundance of satellite halos in gravitational
lens galaxies, and apply our method to a sample of 7 lens systems.  After using
Monte Carlo simulations to verify the method, we find that
substructure comprises $f_{sat}=0.02$ (median, $0.006  < f_{sat} < 0.07$ at 90\% 
confidence) of the mass of typical lens galaxies, in excellent agreement 
with predictions of CDM simulations.  We estimate a characteristic critical radius for the
satellites of $0\farcs0001 < b < 0\farcs006$ (90\% confidence). For
a $dn/dM \propto M^{-1.8}$ ($M_{\rm low} < M < M_{\rm high}$) satellite 
mass function, the critical radius provides an estimate that the 
upper mass limit is $10^6 M_\odot \ltorder M_{\rm high} \ltorder 10^9 M_\odot$.
Our measurement confirms a generic prediction of CDM models, and may 
obviate the need to invoke alternatives to CDM like warm dark matter or
self-interacting dark matter.
\end{abstract}


\section{Introduction}

A discrepancy between the number of satellite halos expected from CDM
simulations and the observed numbers of Galactic satellite galaxies 
is part of the prosecution's case for a crisis in the CDM scenario for
structure formation (e.g. Kauffmann~\cite{Kauffmann93}, Moore et al.~\cite{Moore99},
Klypin et al.~\cite{Klypin99}).  Suggested solutions range from the mundane,
such as the inhibition of star formation in the satellites by
photoionization (e.g. Klypin et al.~\cite{Klypin99},
Bullock, Kravtsov \& Weinberg~\cite{Bullock00}), to the exotic, 
such as the disruption of
the satellites by self-interacting dark matter (e.g. Spergel \& Steinhardt~\cite{Spergel00}) or changes
in the power spectrum (e.g. Bode, Ostriker \& Turok~\cite{Bode01}, Colin, Avila-Reese \& 
Valenzuela~\cite{Colin00}).  The satellite crisis must also be
closely related to the more general problem that the number of low 
luminosity galaxies diverges only as $1/L \sim 1/M$ while the number 
of CDM halos diverges as $\sim 1/M^2$, implying that the probability 
of forming a visible galaxy in a low mass halo must diminish as 
$\sim M$ (e.g. Scoccimarro et al.~\cite{Scoccimarro01},
Kochanek~\cite{Kochanek01b}, Chiu, Gnedin \& Ostriker~\cite{Chiu01}).
In principle, the measured abundances of satellite halos should
provide a strong test of the CDM scenario, but 
because the satellites used as evidence for a problem have low 
luminosities and (in many cases) low surface brightness, it is
difficult to apply this test to any galaxy
besides our own.  Moreover, the test only considers the numbers of 
satellites with detectable optical emission, which is at best
a lower bound on the number of CDM halos.

Gravitational lensing is the only probe which avoids both of these
limitations, as was already noted by Moore et al.~(\cite{Moore99}).
First, the test can be applied to many lens systems
spanning a range of redshifts and physical properties.  Second,
because lensing phenomena couple directly to mass, lenses are
sensitive to both luminous and dark substructures in CDM halos.
Mao \& Schneider~(\cite{Mao98}) pointed out that
the anomalous image flux ratios observed in several lenses, 
particularly B1422+231, could be explained by substructures such
as low mass satellites in the primary lens galaxy.  The 
primary lens magnifies the perturbations from the substructure,
making the brightest images particularly susceptible to the
effects of substructure.  
Recently, Metcalf \& Madau~(\cite{Metcalf01}) quantified the
effects of CDM satellites using simulations and found that
the effects should be readily detected, and Chiba~(\cite{Chiba01})
demonstrated that plausible CDM satellite distributions could
explain the anomalous flux ratios in B1422+231 and PG1115+080.
Detailed studies of B1422+231 (Keeton~\cite{Keeton02},
Bradac et al.~\cite{Bradac02}) find that the observed
perturbations require substructure with mass scales 
comparable to CDM substructure ($\gtorder 10^6M_\odot$) 
rather than stellar microlensing,  and Metcalf \& Zhao~(\cite{Metcalf02})
have shown that the anomolous flux ratios cannot be
reproduced in a large family of smooth potentials for
the primary lens.  

The missing link is an approach for analyzing the gravitational
lens data to estimate the properties of the satellite population.
In this paper we develop such an analysis method and apply it to
a sample of 7 lenses to estimate the surface density and 
characteristic mass of the perturbing satellites.  
We focused on analyzing four-image radio lenses because using
the radio lenses eliminates the problem of dust extinction, and 
minimizes the problems from stellar microlensing due to the relatively 
large source size (see Koopmans \& de Bruyn~\cite{Koopmans00}).
We analyzed the lenses MG0414+0534 (Hewitt et al.~\cite{Hewitt92}), 
B0712+472 (Jackson et al.~\cite{Jackson98}),
PG1115+080 (Weymann et al.~\cite{Weymann80}), B1422+231 (Patnaik et
al.~\cite{Patnaik92}), 
B1608+656 (Fassnacht et al.~\cite{Fassnacht96}),
B1933+503 (Sykes et al.~\cite{Sykes98})  and
B2045+265 (Fassnacht et al.~\cite{Fassnacht99}).
Of these 7 four-image lenses, 6 show anomalous flux ratios
which might be due to the effects of substructure.
We develop our formalism, characterize our model for the
satellite distribution and test our analysis methods
in \S\ref{form}. We apply it to the lens sample
in \S\ref{data}.  In \S\ref{discuss} we review our conclusions and 
their limitations and then outline the observations needed to 
improve them.

\section{Analyzing the Effects of Substructure on Gravitational Lenses}
\label{form}
\def\bfx{{\bf x}}
\def\bfu{{\bf u}}
\def\bfp{{\bf p}}
\def\bfm{{\bf M}}
\def\bfa{{\bf \alpha}}
\def\bfc{{\bf C}}
\def\bfD{{\bf D}}
\def\bfd{{\bf d}}
\def\bfE{{\bf E}}
\def\bfe{{\bf e}}

In this section we outline our mathematical approach to analyzing
the lenses to determine the properties of the substructure (\S2.1)
and the physical model we use for the satellites composing the
substructure (\S2.2).  In \S2.3 we discuss the relationship
between our model and the physical properties of the substructure
such as its fractional surface density, mass and velocity
scales and linear sizes.  In \S2.4 we outline our Monte
Carlo models and test the analysis method.

\subsection{A Linearized Approach to Analyzing Substructure}

Unlike the primary lens galaxy, which we can observe directly to 
determine its position and optical properties (e.g. Lehar et al.~\cite{Lehar00}, 
Kochanek et al.~\cite{Kochanek00}),  we can detect
substructure only through its effects on the positions and 
fluxes of the lensed images.  This means that estimates for
the properties of the substructure will be difficult to 
separate from the properties of the primary lens (the ``macro''
model to adopt the language of the quasar microlensing literature)
because many perturbations will be degenerate with changes
in the macro model.  For this reason, Mao \& Schneider~(\cite{Mao98})
focused on merging image pairs where macro models
must generically predict similar image fluxes but the 
observations sometimes find very different fluxes.
We will instead allow the macro model to compensate for
the effects of substructure as part of our analysis.  If we
confine our analysis to typical four-image (two-image) lenses
we have 14 (8) constraints for determining the 10 parameters
of a realistic macro lens model.  For a four-image lens, the
macro model is overconstrained and we can attempt to estimate
the properties of the substructure.  Because we have typically
found that it is relatively easy to fit image positions and 
hard to fit flux ratios using standard lens models, we expect
the deflection perturbations from substructure to be small or 
degenerate with the parameters of the macro model, and the magnification 
perturbations to be larger and non-degenerate.    

We model the lens by combining 
a macro lensing potential $\phi(\bfx,\bfp)$ defined by a set
of parameters $\bfp$, with a localized perturbing potential
for each image $\delta\phi_i(\bfx)$.  For later notational
simplicity, the source position and flux are considered 
part of the parameter vector $\bfp$.
The time delay surface near image $i$ is  
\begin{equation}
   \tau = { 1 \over 2 } (\bfu-\bfx)^2 - \phi(\bfx,\bfp) 
       - \delta\phi_i(\bfx) = \tau_0(\bfx,\bfp) - \delta\phi_i(\bfx),
\end{equation}
and we find images at solutions of $\nabla\tau =0$ with
an inverse magnification tensor of $\bfm^{-1}=\nabla\nabla\tau$
and magnification $M=|\bfm|$
(see Schneider, Ehlers \& Falco~\cite{Schneider92}). 

We assume the substructure produces small perturbations, so
we can simplify its effects by expanding the lens equations
as a linear perturbations to a macro model for each image $i$.  
We would like to linearize the equations so that the process
of adjusting the macro models to compensate for the effects of
substructure can be done rapidly.
For macro parameters $\bfp_0$ we find images at positions $\bfx_i^{(0)}$
with magnification tensors $\bfm_i^{(0)}$ at the solutions
to $\nabla\tau_0(\bfx_i^{(0)},\bfp_0)=0$.  
Expanding the lens equations 
about these solutions, the perturbed image positions are
\begin{equation}
   \bfx_i^{(1)} = \bfx_i^{(0)} + 
        \bfm_i^{(0)} \cdot \left( \delta\bfx_i - \Delta\bfp \cdot \bfc_i \right)
\label{pertposeqn}
\end{equation}
where $\delta\bfx_i = \nabla \delta\phi_i$ is the deflection
produced by the substructure and 
\begin{equation}
    \bfc_i = { d \nabla\tau_0 \over d\bfp } 
\end{equation}
evaluated at $\bfx=\bfx_i^{(0)}$ and $\bfp=\bfp_0$ 
is the change in the macro model deflections produced by a small
change  $\Delta\bfp=\bfp-\bfp_0$ in the macro model parameters.
The perturbed image magnification is
\begin{equation}
    M^{(1)} = M^{(0)} + { d M \over d \delta\bfm } \cdot \delta\bfm
      +  { d M \over d \bfp} \cdot \Delta\bfp +
         { d M \over d \bfx} \cdot \Delta\bfx
\end{equation} 
which becomes
\begin{equation}
    M^{(1)} = M^{(0)} \left( 1+ \delta m_i + \Delta\bfp \cdot \bfD_i' + \delta\bfx_i \cdot \bfE_i
         \right) 
\label{pertmageqn}
\end{equation}
where 
\begin{equation}
      \delta m_i = 
  -\hbox{tr}\left( \bfm \delta\bfm_i^{-1} \right)
        \quad\hbox{for}\quad \delta\bfm_i^{-1} = -\nabla\nabla \delta\phi_i,
\end{equation}
is the perturbation in the magnification due to the effects of the substructure
on the magnification tensor,
\begin{equation}
   \bfD_i' = { 1 \over M }
     \left( {dM\over d\bfp} - {dM\over d\bfx} \cdot \bfm \cdot \bfc \right)
\end{equation}
is the perturbation due to changing the lens parameters, and
\begin{equation}
   \bfE = { 1 \over M }{ d M \over d \bfx} \cdot  \bfm 
\end{equation}
is the perturbation due to the effects of substructure on the deflections.  
The functions $\delta m_i$, $\bfD_i'$ and $\bfE$ are all evaluated at 
$\bfx=\bfx_i^{(0)}$ and $\bfp=\bfp_0$.
The flux of image $i$ is $f_i = f_s M_i = f_i^{(0)} (1+\Delta f) M_i$, which we 
can linearize as
\begin{equation}
    f_i^{(1)} = f_i^{(0)} \left( 
      1 + \delta m_i + \Delta\bfp \cdot \bfD_i + \delta\bfx_i \cdot \bfE_i 
      \right)
\end{equation}
where $\Delta\bfp \cdot \bfD_i = \Delta\bfp \cdot \bfD_i' + \Delta f$ and
the fractional change in the source flux $\Delta f$ is considered one of 
the model parameters $\bfp$. 

We detect substructure as residuals in fits to the lens parameters which cannot
be modeled by the macro potential.  If we use a $\chi^2$ statistic, the fit
statistic for the image positions is 
\begin{equation}
    \chi^2_a = \sum_{i=1}^N 
      \left( { \bfx_i^{obs} - \bfx_i^{(0)} - \bfm_i^{(0)}(\delta\bfx_i -\Delta\bfp \cdot \bfc_i)
               \over \sigma_a}  \right)^2
\label{chipos}
\end{equation}
where $\sigma_a \simeq 0\farcs003$ is the uncertainty in the observed image positions
$\bfx_i^{obs}$.  
The fit statistic for the image fluxes is
\begin{equation}
    \chi^2_f = \sum_{i=1}^N \left( 
  { f_i^{obs} - f_i^{(0)}(1+\delta m_i +\Delta\bfp \cdot \bfD + \delta\bfx_i \cdot \bfE_i)
         \over \sigma_{f,i} } \right)^2
\label{chiflux}
\end{equation}
for observed fluxes $f_i^{obs}$ and flux uncertainties $\sigma_{f,i} \simeq 0.05 f_i^{obs}$.
The lens position, if measured, is constrained by 
$\chi^2_l = (\bfx_l^{obs}-\bfx_l^{(0)}-\Delta\bfp_l)^2/\sigma_l^2$ where 
$\Delta\bfp_l$ represents the perturbations to the lens position and 
$\sigma_l \simeq 0\farcs003$ is the uncertainty in the observed position
$\bfx_l^{obs}$.  Because all the terms entering the fit statistic depend
only linearly on the $\Delta \bfp$, the fit statistic is a quadratic of the form 
$\chi^2 = \chi^2_0 + 2\Delta\bfp \cdot { \bf I} + \Delta\bfp \cdot {\bf J }\cdot  \Delta\bfp$
which is minimized for $\Delta\bfp = - {\bf J }^{-1} \cdot { \bf I}$  with value
$\chi^2_{min} = \chi^2_0 - { \bf I} \cdot {\bf J }^{-1}\cdot  { \bf I}$. If the macro
model were held fixed, the goodness of fit would be $\chi^2_0$, combining
the effects of the substructure with the differences between the observations
and the initial macro model.  After adjusting the macro model, the
correction $\Delta\chi^2={ \bf I}\cdot  {\bf J }^{-1} \cdot { \bf I}$ represents the ability
of the macro model to mimic the effects of substructure.

\def\bfs{{\bf s}}
In order to test the CDM predictions for substructure we need to estimate the 
statistical properties of the substructure rather than discuss the evidence
for perturbations from substructure in individual lenses.  For a statistical
model of the substructure characterized by parameters $\bfs$, the probability
of the $\alpha$th substructure realization for lens $j$, $\delta_{\alpha j}$,
is $P(\delta_{\alpha j}|\bfs)$.  Given a concrete model for the substructure
$\delta_{\alpha j}$ we can compute the likelihood for fitting the data as the
likelihood of obtaining the $\chi^2$ statistic we find after reoptimizing
the macro lens model given the substructure realization, $P(D_j|\delta_{\alpha j})$.
The Bayesian probability of the model parameters given the data for
$j=1\cdots N$ lenses is
\begin{equation}
    P(\bfs,\delta_{\alpha j}| D_j) \propto 
     P(\bfs) \Pi_{j=1}^N P(D_j|\delta_{\alpha j})P(\delta_{\alpha j}|\bfs)
\label{bayeseq0}
\end{equation}
where $P(\bfs)$ is the prior probability distribution of the $\bfs$. As in
all Bayesian probabilities, the expression summed over all variables is
normalized to unity.  
In estimating the statistical properties of the substructure, 
we are not interested in the likelihoods of the individual realizations,
but in the marginalized distribution 
\begin{equation}
    P(\bfs| D_j) \propto P(\bfs) \Pi_{j=1}^N 
       \sum_{\alpha=1}^M P(D_j|\delta_{\alpha j})P(\delta_{\alpha j}|\bfs)
\label{bayeseq}
\end{equation}
where we sum over the $\alpha =1\cdots M$ Monte Carlo realizations of the substructure
for each lens.  Typically we used $M=10^5$ realizations.  As we vary the 
statistical properties of the substructure $\bfs$, the fraction of the 
realizations $\delta_{\alpha j}$ which significantly improve the goodness of
fit varies.  These changes in the fraction of realizations which improve
the fit relative to the macro model alone allow us to estimate the
parameters $\bfs$ describing the substructure.  

\def\bk{{\bf k}}
\def\bv{{\bf V}}
In our final analysis we used random realizations of perturbing satellites
to estimate the perturbations.  It is worth mentioning, however, that the
problem can be fully linearized if we use a Gaussian model for the
perturbations.  If $\bk=\lbrace\delta m_i,\delta\bfx_i\rbrace$ is
a $d$-dimensional vector of the perturbation variables, and they have a covariance
matrix $\bv^{-1}=\langle \bk^T \bk\rangle$, then the Gaussian model
for the probability distribution of the perturbations is
\begin{equation}
    P(\bk) = |\bv|^{1/2} (2\pi)^{-d/2}
       \exp\left( -{1\over 2} \bk \cdot \bv \cdot \bk \right).
\label{pertprob}
\end{equation}
The matrix $\bv$ is proportional to the inverse of the satellite
surface density, so by combining eqns.~(\ref{chipos}), (\ref{chiflux})
(\ref{bayeseq}) and (\ref{pertprob}), the marginalizing integrals over the shifts
in the lens model and the distribution of satellite realizations
can be done analytically using standard methods for linear algebra
and Gaussian integrals to leave an expression depending only on
the statistical properties of the substructure.  We did not use
this for our actual analysis because it was relatively easy to
perform the necessary Monte Carlo realizations and integrals 
needed to reproduce the true probability distribution for the
substructure and its correlations, but we did use it as an 
internal check on our results.  We mention it here because other
studies may find it to be of similar utility.

\subsection{Substructure Models }

Before applying the above formalism to observed lens systems, it is
necessary to calculate the expected amplitude of perturbations to 
lensed images from CDM substructure.
This involves calculating the expected levels of
astrometric shifts in the image positions, and the RMS fluctuations in
the local convergence and shear.  These quantities are then 
magnified by the local magnification tensor of the smooth macro-model.

We model CDM substructure by randomly laying down subclumps of surface
density.
The substructures seen in CDM simulations appear to have mass profiles
consistent with the `universal' NFW profile, however for simplicity we
will treat them as pseudo-Jaffe models (density $\rho \propto r^{-2}(r^2+a^2)^{-1}$,
see Munoz, Kochanek \& Keeton~\cite{Munoz01}), 
with convergence, the surface density in units of the critical surface density,
\begin{equation}
\kappa(r) = {\Sigma \over \Sigma_c} = 
  { b \over 2 } \left[ { 1 \over r } - { 1 \over (r^2+a^2)^{1/2} } \right]
\end{equation}
where the critical surface density for lensing is
$\Sigma_c=c^2 D_{OS} /4\pi G D_{OL}D_{LS}$.  
Here, $b$ is a length scale similar to the Einstein radius of the
subclump, and $a$ is a tidal or break radius.  Note that 
$b$, $r$ and $a$ are angular lengths, which are related to
physical sizes by multiplication by the distance to the lens $D_{OL}$.  
The total mass of a clump is $M=\pi b a\Sigma_c$ where $\Sigma_c$
is the critical density in angular units.  If the surface mass
density of the perturbers is $\Sigma$, the number of perturbers
per unit area is $N=(\Sigma/\Sigma_c)/\pi b a$.  To leading order,
the variance in the image deflection, convergence and shear are
\begin{equation}
  \langle | \delta \bfx|^2\rangle \simeq {3\over2} { \Sigma \over \Sigma_c } ba
  \qquad
  \langle \kappa^2 \rangle \simeq \langle \gamma^2 \rangle \simeq   
   { 1 \over 2} { \Sigma \over \Sigma_c } { b \over a } \ln { a \over s}
\label{varx}
\end{equation}
where we must introduce a core radius $s$ to make the variance in
the convergence and shear finite.  These perturbations may then be
magnified by the macro model to produce the observed perturbations
in the image positions and fluxes.  
If the scale $a$ is determined
by the satellite's tidal radius, we have
$a=(b b_0)^{1/2}$ where $b_0$ is the critical radius of the 
primary lens (see e.g. Metcalf \& Madau~\cite{Metcalf01}).  Thus, for
a fixed satellite surface density, the variance of the astrometry
perturbation in units of the critical radius of the macro lens
is roughly
\begin{equation}
  { \langle | \delta \bfx|^2\rangle^{1/2} \over b_0 } \simeq
     10^{-3} \left( { 10 \Sigma \over \Sigma_c } \right)^{1/2} 
             \left( { 10^3 b \over b_0 } \right)^{3/4},
\label{covar1}
\end{equation}
while the variance in the shear and convergence is roughly
\begin{equation}
\langle \kappa^2 \rangle^{1/2} \simeq \langle \gamma^2 \rangle^{1/2}
   \simeq 0.13 \left( { 10 \Sigma \over \Sigma_c } \right)^{1/2}
        \left( { 10^3 b \over b_0 } \right)^{1/4} 
        \left( { \ln \Lambda \over 10 }\right)^{1/2}
\label{covar2}
\end{equation}
where $\ln \Lambda = \ln (\sqrt{b b_0}/s) \sim 10$.  

\subsection{Physical Scales and Interpretations}

For this first attempt at modeling substructure, we consider satellites with
constant surface density $\Sigma/\Sigma_c$ and critical radius $b$.  We scale
the satellite break radius like a tidal radius with $a=(b b_0)^{1/2}$ for
$b_0\equiv 1\farcs0$.   Near the critical radius of a moderately elliptical
isothermal lens the surface density is $\Sigma_c/2$, so the projected
satellite mass fraction is $f_{sat}=2\Sigma/\Sigma_c$.  In the cylinder
(sphere) defined by the Einstein radius, roughly 50\% (10\%) of the mass
is dark matter (see, e.g., Keeton~\cite{Keeton01}).  The Einstein ring,
where we make the measurement, is typically 1.0-1.5 effective radii from
the lens center and the dark matter fraction will be significantly higher
than these average values at the edge of the cylinder where we see
the images.  Hence, we can interpret
$f_{sat}$ as the fraction of the dark matter near the Einstein ring 
in substructure with only modest baryonic corrections.

We use only observational parameters in our models, which means that
the physical parameters will have small shifts between the lenses we
consider because of the changing lens and source distances. To provide
a sense of the physical scales, consider the parameters for  PG1115+080
with source and lens redshifts of $z_s=1.72$ and $z_l=0.31$ respectively.
The inner circular velocity of a satellite is 
$v_{circ}=9.7(b/0\farcs001)^{1/2}$~km/s,
and its mass is $M=3.4 (b/0\farcs001)^{3/2} \times 10^6 h^{-1} M_\odot$.
The angles $b$ and $a$ correspond
to proper distances of $3(b/0\farcs001)h^{-1}$~pc and
$100(b/0\farcs001)^{1/2}h^{-1}$~pc compared to the average
distance of 1\farcs16 or $3.6h^{-1}$~kpc of the images from the
lens center.  For the other lenses, the distances scale linearly
with $D_{OL}$, the circular velocity scales as $(D_{LS}/D_{OS})^{-1/2}$
and the mass scales as $D_{OS}/D_{OL} D_{LS}$.  These changes
between lenses are sufficiently small compared to our logarithmic
uncertainties to ignore.

We will use satellites with fixed properties in our models, so our
estimate of the mass scale is a weighted average of the satellite
masses.  Given our statistical uncertainties models with a mass
spectrum are unwarranted, and we should be able to estimate the
effects of using a mass spectrum simply by matching the variance
in the shear and astrometry perturbations (Eqn. \ref{varx}).  
The mass function of the satellite halos is $dN/dM \propto M^{-\alpha}$
with $1.7 \ltorder \alpha \ltorder 1.8$
(e.g.  Moore et al.~\cite{Moore99}, Klypin et al.~\cite{Klypin99},
Metcalf \& Madau~\cite{Metcalf01}, Springel et al.~\cite{Springel01},
Helmi et al.~\cite{Helmi02}) which we limit to a finite
range $M_{\rm low} < M < M_{\rm high}$ to avoid divergences
in the total mass.  With $\alpha <2$, only the upper mass
limit is important in estimating the perturbations.  Our
effective substructure mass $M$ is related to the upper
mass limit by $M= M_{\rm high}(2-\alpha)/(3-\alpha)=M_{\rm high}/6$
if we match the amplitude of the astrometry perturbations
and by $M= M_{\rm high}((2-\alpha)/(7/3-\alpha))^3=M_{\rm high}/20$
if we match the shear perturbations.
Given the precision with which we can currently estimate the
characteristic mass scale, we choose not to include a satellite
mass function.  Crudely, we can estimate that
$M_{\rm high} \sim 10-20 M$. 
\footnote{For $\alpha=2$ the results become logarithmically
sensitive to the lower mass limit.  If we match the astrometric
perturbations, our mass scale corresponds to 
$M = M_{\rm high}/\ln C \simeq M_{\rm high}/10$ where
$C=M_{\rm high}/M_{\rm low}$ is  the ratio of the
upper and lower mass limits.  If we match the 
shear perturbations, assuming the Coulomb logarithm is
held fixed, our mass scale corresponds to 
$M = M_{\rm high}(3/\ln C)^3 \sim M_{\rm high}/30$.}

\begin{figure}
\centerline{\psfig{figure=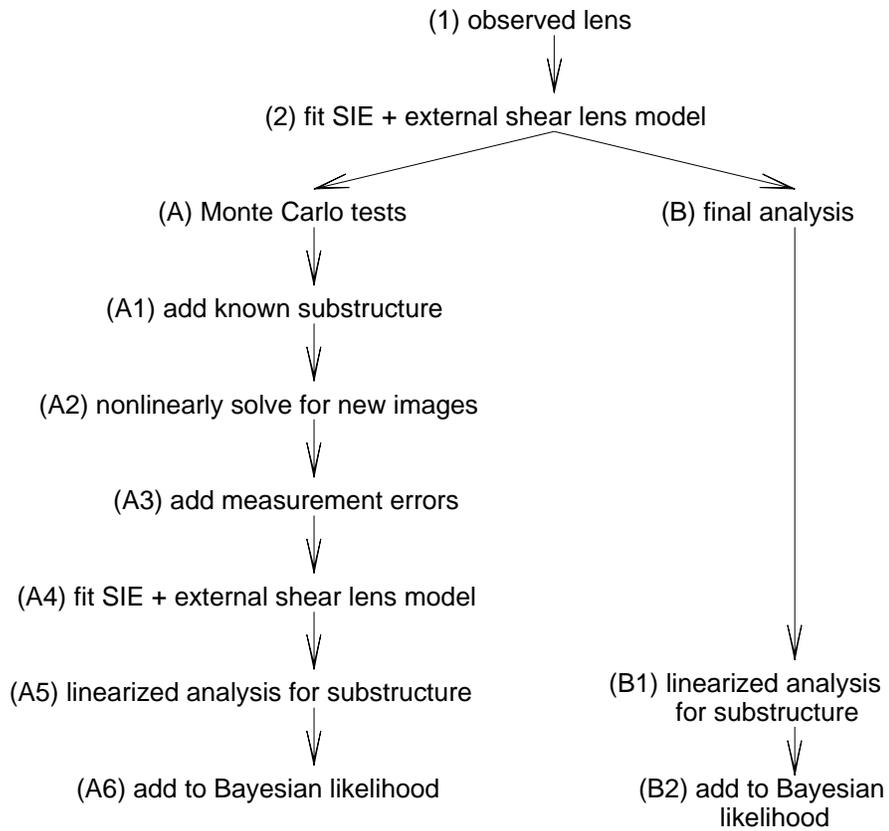,width=6.0in}}
\caption{ Outline of analysis procedures.
  \label{fig:process}
  }
\end{figure}

\subsection{Analysis Procedures and Monte Carlo Tests}

In Fig.~\ref{fig:process} we outline our procedures for 
analyzing the lens data and for testing our method to
ensure that it can recover the properties of the 
substructure accurately by following the treatment we
use for each lens in the sample.    
We start (\#1) with the available data on the lens.  The
first processing step (\#2) is to model the lens with the 
{\it lensmodel} package (Keeton~\cite{gravlens}) using 
singular isothermal ellipsoids (SIE) in external shear
fields for the mass distribution of the primary lens galaxy.
The SIE model is the only standard lens model which is  
consistent with general properties of the lens sample (see,
e.g. Cohn et al.~(\cite{Cohn00}), and references therein).  
Where needed, we added additional SIE lens components so as to
reproduce the best standard models for each system.   
We used the observed astrometric uncertainties but 
broadened the uncertainties in the flux ratios to 
20\% to compensate for systematic errors and the
contaminating effects of the substructure on the
flux ratios.  Effectively we follow the procedures
suggested by Mao \& Schneider~(\cite{Mao98}) for
modeling lenses in the presence of substructure.

For analyzing the real data (steps B1-B2 in Fig.~\ref{fig:process}) we apply our 
linearized analysis method from \S2.1 to each lens using 
the best fit macro model from step \#2 as the reference 
model ($\bfp_0$) supplying the reference image positions
and magnifications ($\bfx_i^{(0)}$, $\bfm_i^{(0)})$.
Given the mass scale $b$ and surface
number density $N$ of the substructure we determine the angular
radius  $r_n = (n/N)^{1/2}=(n b a \Sigma_{crit}/\Sigma)^{1/2}$
inside which we expect to find $n$ perturbing satellites.
For each realization of the substructure (the $\delta_{\alpha j}$
in Eqns.~(\ref{bayeseq0}) and (\ref{bayeseq})) we added $n=10$
perturbing satellites inside radius $r_{10}$ from each image with 
corrections to avoid over counting in models where the $r_{10}$ 
regions of the individual images overlap.  Each satellite perturbed all
images, an effect which becomes important for mass scales
$b \gtorder 0\farcs01$. The model is in some senses still
a ``local'' approximation because we assume a constant surface
density near all images and we do not generate a full, global
realization of the substructure distribution. The more distant
satellites produce perturbations which are difficult to distinguish
from changes in the macro model.  We varied only the mass
scale $b$ and surface density $\Sigma/\Sigma_c$ of the substructure
using logarithmic priors for the two variables ($P(b) \propto 1/b$ and
$P(\Sigma) \propto 1/\Sigma$). The tidal
radius was always set to $a=(b b_0)^{1/2}$ with $b_0\equiv 1\farcs0$.
For each value of the mass scale and the surface density, we generated
$10^5$ random realizations of the substructure.  
{\it All parameters of the macro model are reoptimized for every
substructure realization} by minimizing the fit statistics in
Eqns.~(\ref{chipos}) and (\ref{chiflux}) combined with any
ancillary constraints like the position of the primary lens
galaxy.  The Bayesian likelihood distribution (Eqn.~\ref{bayeseq})
is constructed by combining the likelihoods of fitting the data for lens
$j$, $P(D_j|\delta_{\alpha j})$ for each of the $\alpha=1\cdots 10^5$ 
substructure realizations made for each of the set of substructure 
parameters $\bfs$.

We tested the algorithm using Monte Carlo models following the
steps A1-6 in Fig.~\ref{fig:process}.  The objective of the 
Monte Carlo sequence is to start from the best fit macro 
model of each lens found in step \#2 and then by adding
substructure and noise generate a synthetic set of lens
data which should be analogous to the real data.  We start
by taking the best fit model for the lens (from \#2) and 
using its parameters and source properties as the true
properties of a new Monte Carlo model.  
In step A1 we randomly place $n=5$ perturbing satellites
inside the radius $r_5$ of each image based on the
desired physical properties (mass scale $b$, surface
density $\Sigma$, tidal radius $a$) of the substructure.  

In step A2 we use the {\it lensmodel} package (Keeton~\cite{gravlens}) 
to find the non-linear solutions for the new image positions and 
fluxes including all the substructure but keeping the macro 
model and source properties fixed.  We used $n=5$ perturbing
satellites per image because of limitations on the maximum
number of lenses in {\it lensmodel}.  Tests varying the
number of perturbers used both to generate and analyze
the data suggested that the choices have no effects on 
the results in the sense that any biases are small compared
to the statistical uncertainties. A modest fraction
of realizations for B2045+265 produced extra
images.  We discarded these realizations.  The existence
of these solutions suggests that the substructure profile 
shape may be constrained by the production of extra images,
but an exploration of these additional parameters is 
beyond the scope of our current study.  After adding
measurement errors to the image positions, lens positions
and image fluxes in step A3, we have a set of synthetic
lens data that should be a realistic Monte Carlo 
model of the real data we started with in 
step \#1.  The remainder of the analysis is identical to that 
for the real data.  Step A4 matches step \#2 where we fit
the synthetic data using only a smooth macro model to
provide the reference models and images for performing
the substructure analysis.  
Step A5 matches step B1 for the real data, where we
apply our linearized substructure analysis to the
noisy synthetic data and the reference model, and
step A6 matches step B2 where we combine the results
to estimate the Bayesian likelihood distributions
for the substructure parameters. 

\begin{figure}
\centerline{\psfig{figure=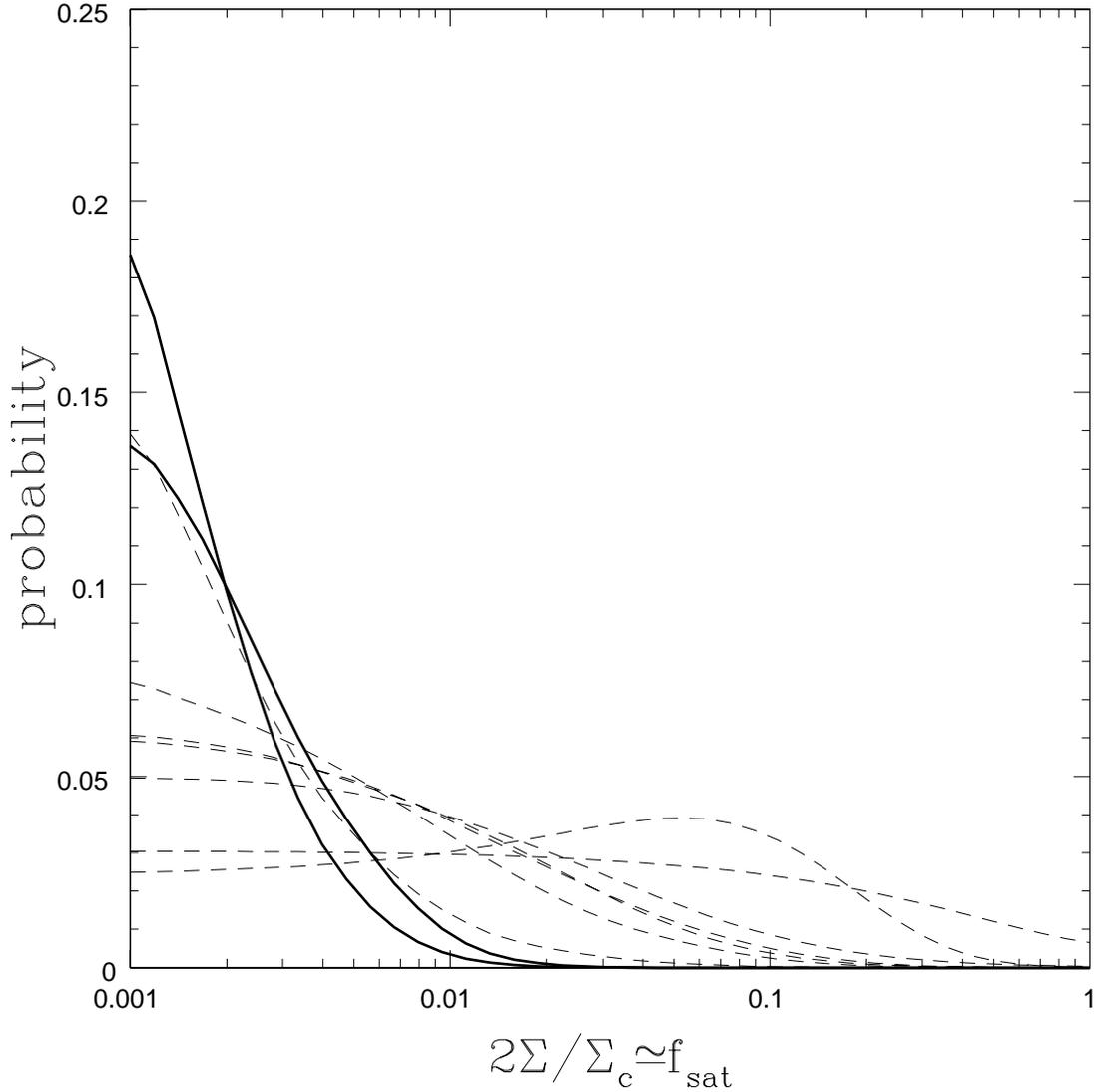,width=6.0in}}
\caption{ Two Monte Carlo simulations of the effects of measurement error.  
  The dashed curves show the likelihood distribution for each lens in the
  first realization and the heavy solid curves show the final Bayesian
  probability distributions for the two realizations of 7 lenses. 
  Our formal upper limit is $f_{sat} \ltorder 0.004$ in both trials,
  but the limit would be lower had we extended the calculation 
  beyond the range $10^{-3} < f_{sat} < 1$.
  \label{fig:montenull}
  }
\end{figure}

\begin{figure}
\centerline{\psfig{figure=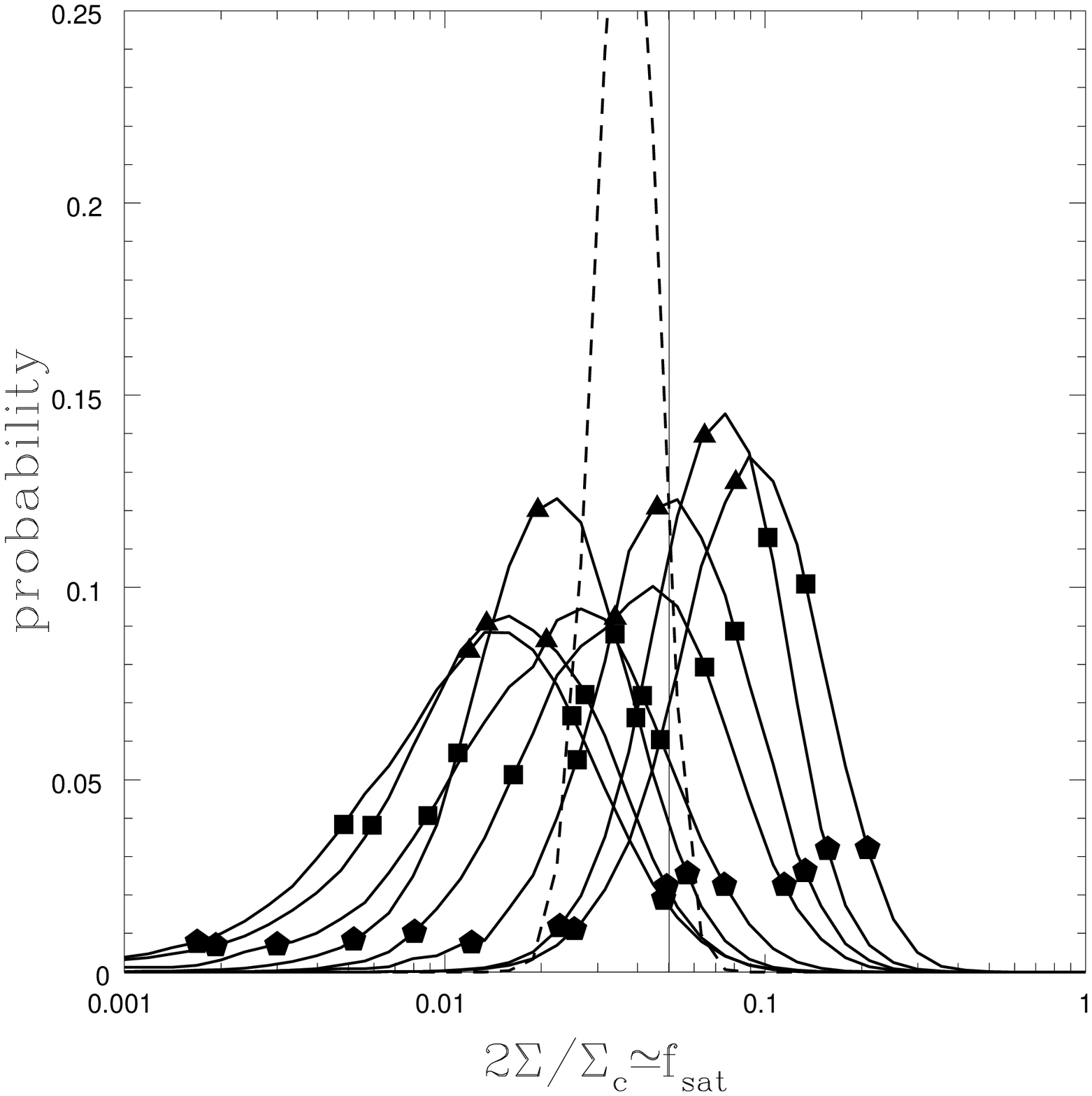,width=6.0in}}
\caption{ Eight Monte Carlo simulations with $f_{sat}=0.05$ and $b=0\farcs001$.  
  The light dashed curves show the likelihood distribution for each lens in the
  first realization, the heavy solid curves show the final Bayesian probability 
  distributions for all eight realizations of 7 lenses each, and the heavy
  dashed curve shows the combined probability of all 8 realizations. This
  latter case mimics a sample of 56 lenses.  The points on the heavy curves 
  mark the median probability (triangles) and the regions encompassing 68.3\% 
  (1$\sigma$, squares), and 95.4\% (2$\sigma$, pentagons)
  of the likelihood in the Bayesian probability distribution.  The vertical
  line marks the true value of $f_{sat}=0.05$.
  \label{fig:montesat}
  }
\end{figure}

We illustrate the ability of our linearized analysis method to correctly
extract the properties of substructure in three steps.
First, we examined the sensitivity
of our surface density estimate $f_{sat}$ to measurement errors. Next,
we tested our ability to estimate $f_{sat}$ when the satellite masses
and structures were fixed to their true values.  Finally, we tested
our ability to estimate simultaneously the surface density $f_{sat}$
and the mass scale $b$.  In each case we generate a Monte Carlo
data set consisting of a perturbed realization for each of the
7 lenses in our sample and then analyze it using the same
procedures we apply to the real data.

In our first test we examine whether measurement errors can be
mistaken for substructure by adding random astrometry and flux
errors to the models, fitting new macro models, and then analyzing
the synthetic data using our method.  This is unlikely to be a
serious problem for the real data because the best fit models
typically have $\chi^2\gg N_{\rm dof}$ when we use realistic
uncertainties for the image fluxes.  However, this will determine
the range of $f_{sat}$ to which we are sensitive, where the
relevant scales are $10^{-4} \ltorder f_{sat} \ltorder 10^{-3}$
for normal satellite populations and $0.02 \ltorder f_{sat} \ltorder 0.15$
for CDM substructure.  The results of two such simulations are
shown in Fig.~\ref{fig:montenull}. The formal, one-sided 90\%
confidence upper bounds are $f_{sat} \ltorder 0.004$, although
this is very conservative because we only calculated the 
probability over the range $10^{-3} \leq f_{sat} \leq 1$.  The
peak probability and most of the integrated probability comes
from still lower satellite fractions. Lenses with highly
magnified images are more sensitive to substructure and constrain
$\Sigma/\Sigma_c$ more strongly, with the upper limit varying as the
inverse square of the maximum image magnification.  Our lens
sample has two ``low-magnification'' lenses ($M_{max}\ltorder 5$, 
B1608+656, B1933+503) and five ``high-magnification'' lenses
($M_{max}\gtorder 5$, MG0414+0534,  B0712+472, PG1115+080, B1422+231, and 
B2045+265). Individual lenses can even show probability peaks
at larger surface densities, but without the contrast between
the peak and the probability at lower $f_{sat}$ needed to 
produce a signal at higher surface density in the full sample.
If, however, we underestimate the flux errors, 
we can make a spurious detection of the substructure. 
When we examine a range of satellite mass scales $b$ as well as the surface
density, we find that measurement errors produce no preferred mass
scale for the substructure.  

In the second set
of simulations we added perturbing satellites near each image with a
surface density of $f_{sat}=2\Sigma/\Sigma_c = 0.05$ and a mass profile 
defined by $b=0\farcs001$ and $a=0\farcs032$.  The radius encompassing
an average of 5 satellites, $r_5=0\farcs080$, is much smaller than the 
distances between the images.  We experimented with other
models, but the results we present were typical.  
As we show in Fig.~\ref{fig:montesat}, the estimated surface
density is consistent with the surface density used to generate the
data.  In the eight simulations shown in Fig.~\ref{fig:montesat},
the surface density corresponding to the median probability ranges
from $f_{sat}=0.01$ to $0.08$ with uncertainties of a factor
of $2.5$ at 1$\sigma$ and $3.0$ at 90\% confidence.
The true surface density is within the 68.3\% (1$\sigma$) confidence
region in 4 of the 8 simulations and within the 90\% confidence
region for 6 of the 8 simulations.  If we combine all 8 simulations to mimic
a sample of 56 lenses, the surface density estimated by the median
of the Bayesian likelihood distribution is $f_{sat}=0.034$ with 
a 90\% confidence range of $0.023 \leq f_{sat} \leq 0.048$ that 
marginally excludes the true value.  The slightly low value for $f_{sat}$ could
be due to chance, discarding the cases producing additional images, 
linearizing the problem or the local approximation for the
substructure. 

\begin{figure}
\centerline{\psfig{figure=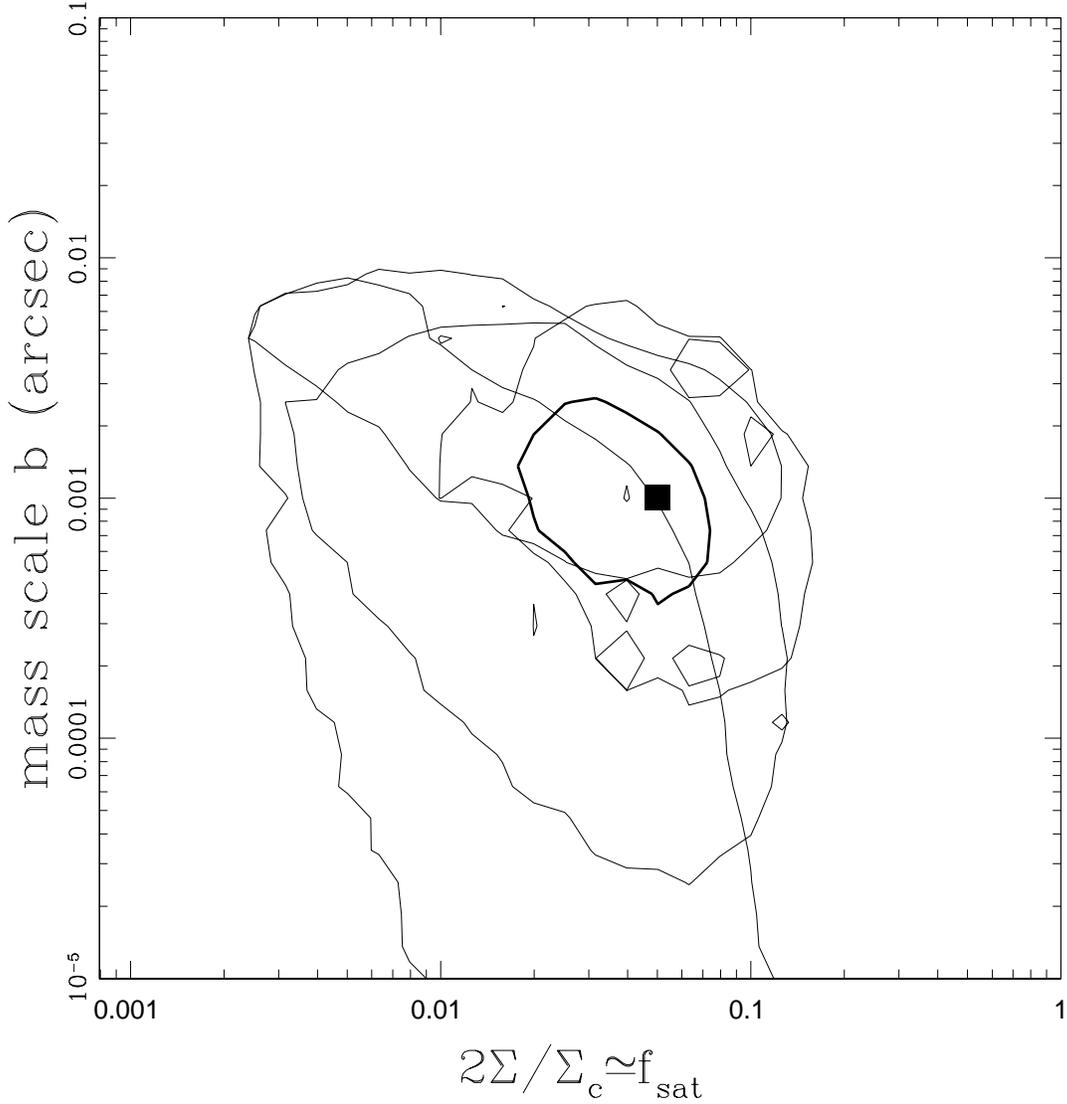,width=6.0in}}
\caption{ Monte Carlo simulations including both the surface density
  $f_{sat}$ and the mass scale $b$.  We show the probability contour
  enclosing 90\% of the total probability for four of the eight 
  models shown in Fig.~\protect{\ref{fig:montesat}}.  The heavy contour 
  shows the result after combining all eight realizations and
  the solid square marks the true model parameters. 
  \label{fig:montesat2d}
  }
\end{figure}

We also examined the likelihood distribution in the two-dimensional 
space of the surface density $f_{sat}$ and the mass scale $b$, holding
the internal structure of the satellites fixed with $a=(b b_0)^{1/2}$
for $b_0\equiv 1\farcs0$.  Fig.~\ref{fig:montesat2d} shows that the
method can recover the mass scale, but less robustly than the surface
density of the satellites.  In these two-dimensional models we find
median estimates for the surface density ranging from $f_{sat}=0.014$
to $0.074$ with a factor of $3.8$ uncertainty at 90\% confidence.
The ability to adjust the mass scale significantly increases the
uncertainty in the surface density.  The median estimates for the
mass scale range from $b=0\farcs00016$ to $0\farcs0027$. The
uncertainty in the mass scale is usually an order of magnitude
at 90\% confidence.  As we discuss in \S2.2, we expect to be more
sensitive to the surface mass density than to the mass scale.
For constant astrometry perturbations we expect
$b \propto f_{sat}^{-2/3}$ and for constant shear or convergence
perturbations we expect $b \propto f_{sat}^{-2}$ (see eqns.
\ref{covar1} and \ref{covar2}).  Neither slope is clearly
reflected in the likelihood contours of Fig.~\ref{fig:montesat2d},
suggesting that both types of perturbations contribute.
If we combine all 8 realizations to mimic a sample of 56 lenses,
we recover the input model with modest uncertainties.  

\begin{figure}
\centerline{\psfig{figure=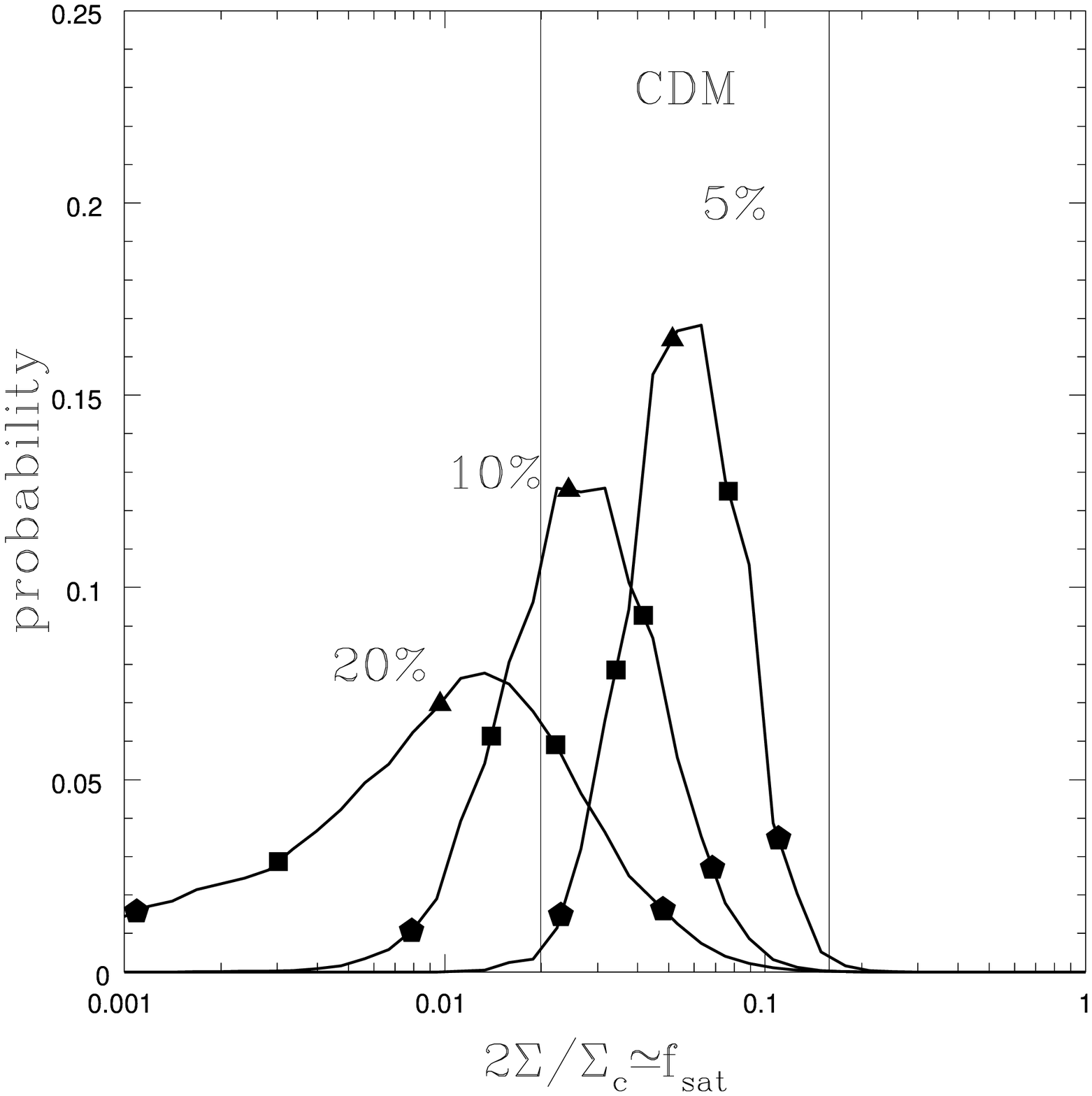,width=6.0in}}
\caption{ Results for the observed lens sample with $b=0\farcs001$.  The
  heavy solid lines show the probability distributions assuming errors
  in the flux ratios of 5\%, 10\% and 20\%.  The points on the curves
  mark the median surface density (triangles) and the regions encompassing 
  68.3\% (1$\sigma$, squares), and 95.4\% (2$\sigma$, pentagons) of the probability.  The
  dashed curves show the contributions from the individual lenses for
  the 10\% case.  The region between the vertical lines is the range
  of substructure mass fractions found in the
  Klypin et al.~(\protect{\cite{Klypin99}}) simulations.
  Normal satellite populations, with $10^{-4} \ltorder f_{sat}\ltorder 10^{-3}$,
  correspond to a region off the left edge of the figure.
  \label{fig:datasat}
  }
\end{figure}
\begin{figure}
\centerline{\psfig{figure=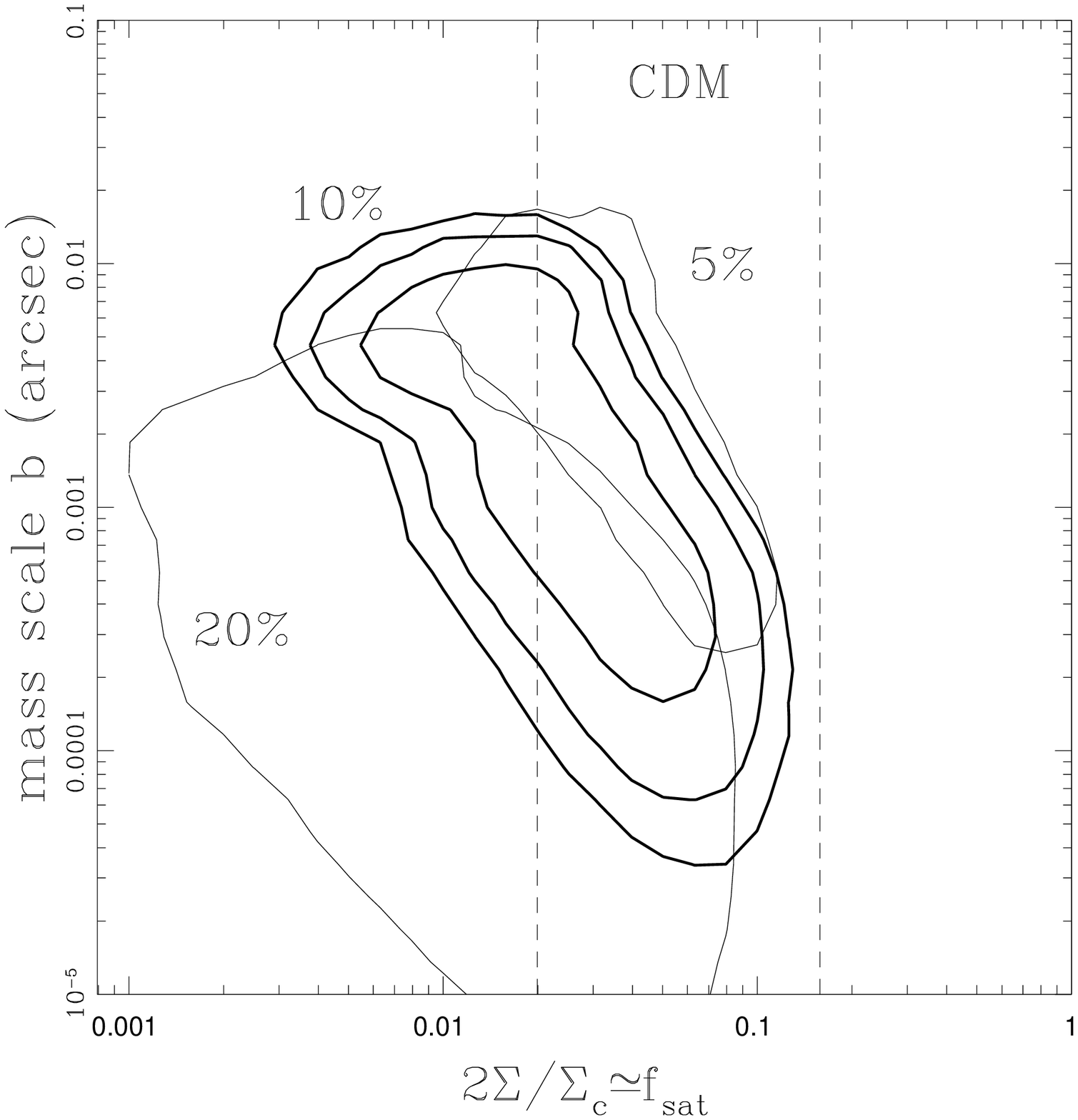,width=6.0in}}
\caption{ Results for the observed lens sample as a function of both the
  surface density $f_{sat}$ and the mass scale $b$.  For the case with
  10\% flux errors we show the probability contours enclosing 68\% 
  (1$\sigma$), 90\%, and 95\% (2$\sigma$) of the total probability 
  using heavy solid curves.  For the cases with 5\% and 20\% flux
  errors we show only the probability contour encompassing 90\%
  of the probability using a light solid curve.  The region between the 
  vertical lines is the range of substructure mass fractions found in the
  Klypin et al.~(\protect{\cite{Klypin99}}) simulations.
  Normal satellite populations, with $10^{-4} \ltorder f_{sat}\ltorder 10^{-3}$,
  correspond to a region off the left edge of the figure.
  \label{fig:datasat2d}
  }
\end{figure}

In summary, the lenses are sensitive to surface densities of substructure
exceeding $f_{sat} \gtorder 0.004$ and samples of 7 lenses can be used to determine
the surface density and mass scale with reasonable accuracy.  Our method
shows no signs of biases in the recovered parameters given the statistical
uncertainties expected for 7 lenses. In much larger lens samples, or
samples with very accurately measured image fluxes, we may underestimate
the surface density slightly as a consequence of the simplifications 
made to allow rapid calculation (linearizing the problem and the ``local''
approximation for the substructure) and the elimination of realizations
generating extra images.

\section{The Properties of Halo Substructure}
\label{data}

Given these limitations we now apply the analysis to the sample of 7
real lenses.  We will assume an average uncertainty in the flux ratio
measurements of 10\%, but report results for uncertainties of 5\% and
20\% as well.  For the few cases where the flux measurement errors are
larger we use the measurement errors instead, but in most cases the
flux measurement errors are dominated by systematic uncertainties rather
than measurement errors.  Amongst the systematic issues are variability
and time delays, wavelength dependencies to the flux ratios, and any
contribution from stellar microlensing.  A detailed examination of these
problems is beyond the scope of our present study.  Given the current
data for most lenses, the image flux uncertainties are certainly lower 
than 20\%, probably lower than 10\% and unlikely to be lower than 5\%.
The errors in the image and lens positions are dominated by measurement 
errors rather than systematic errors.

We first analyzed the data assuming a fixed mass scale of $b=0\farcs001$
and a tidal radius of $a=0\farcs032$.  As shown in Fig.~\ref{fig:datasat},
the results for the real lens sample have qualitative properties that are
very similar to the results of the Monte Carlo simulations shown in
Fig.~\ref{fig:montesat}.  
The median estimate for the surface density
depends on the assume level of systematic uncertainties in the image
flux ratios, with $f_{sat}\equiv 2\Sigma/\Sigma_{crit}=0.051$, $0.024$
and $0.0097$ for flux ratio uncertainties of 5\%, 10\% and 20\% respectively.  
The 90\% confidence ranges for the three cases are $0.027  < f_{sat} < 0.096$,
$0.0098  < f_{sat} < 0.058$ and  $0.0014 < f_{sat} < 0.037$ respectively.
In all three cases the distributions are broadly 
consistent with the $0.02 < f_{sat} < 0.15$ range found in the
Klypin et al.~(\cite{Klypin99}) simulations, and well above the
$10^{-4}\ltorder f_{sat} \ltorder 10^{-3}$ found in visible satellites (see 
Mao \& Schneider~\cite{Mao98}, Chiba~\cite{Chiba01}).  

We also calculated the probabilities as a function of both $f_{sat}$ and
the mass scale $b$ as shown in Fig.~\ref{fig:datasat2d}. With 10\%
flux errors the median estimates for the surface density and mass scale 
are $f_{sat}=0.020$ and $0\farcs0013$ 
with 90\% confidence regions of
$0.0058 < f_{sat} < 0.068$ and $0\farcs0001 < b < 0\farcs007$.    
For a $dn/dM \propto 1/M^2$ ($M_{\rm low} < M < M_{\rm high}$) 
satellite mass function this implies that the upper mass 
scale is in a range
$10^6 M_\odot \ltorder M_{\rm high} \ltorder 10^9 M_\odot$ 
that is consistent with the expectations for satellites.
There is a relatively strong covariance between the parameters
$b$ and $f_{sat}$, with low surface densities requiring higher
mass scales.  The slope of the likelihood contours is very close to 
the $b \propto f_{sat}^{-2}$ slope corresponding to constant shear or 
convergence perturbations (see eqn. \ref{covar2}) rather than the 
flatter $b\propto f_{sat}^{-2/3}$ slope corresponding to constant 
astrometry perturbations (see eqn. \ref{covar1}).
If we assume 5\% flux errors, then the surface
density and mass scales are restricted to larger values,
with $0.013 \ltorder f_{sat} \ltorder 0.078$ and 
$0\farcs00036 \ltorder b \ltorder 0\farcs013$.  If we assume
20\% flux errors, a broader range is permitted, with
$0.0016 \ltorder f_{sat} \ltorder 0.051$ and
$0\farcs000015 \ltorder b \ltorder 0\farcs0023$.  These calculations
neglect the smoothing effects of the finite source size 
($\Delta\theta\sim 0.01$--$1.0$~milliarcsecond), which will wash out the 
perturbations from smaller satellites if $\Delta\theta \gtorder b$. 
With a finite source size, we would require a larger satellite 
fraction to produce the same perturbations to the images.
On a final, if qualitative note, the general properties of the likelihood
distributions for the real data are remarkably similar to those of the Monte 
Carlo simulations.

\section{Discussion}
\label{discuss}

CDM simulations generically produce halos in which $\sim$2-15\% of
the mass is comprised by substructure, which is 50-100 times more mass
than is observed in the satellites of the Local Group (e.g. 
Moore et al.~\cite{Moore99}, Klypin et al.~\cite{Klypin99}). 
This substructure problem, and possible conflicts between rotation
curves and density cusps and the observed and predicted angular
momentum distributions in spiral galaxies have been interpreted
as requiring significant modifications to the CDM paradigm
(e.g. Spergel \& Steinhardt~\cite{Spergel00}, Bode et al.~\cite{Bode01}, 
Colin et al~\cite{Colin00}).

Here we show that the anomalous flux ratios observed in a sample of 7
gravitational lenses can be interpreted as requiring a mass fraction
of $0.006  < f_{sat} < 0.07$ (90\% confidence) in satellite halos that
is remarkably consistent with the CDM predictions.  This estimate 
assumed 10\% errors (measurement $+$ systematic) 
in the estimates of image fluxes, but
the predicted surface density remains consistent with the expectations
for CDM over the plausible 5-20\% range for these uncertainties.  The
estimates are always well above the 
$10^{-4}\ltorder f_{sat} \ltorder 10^{-3}$ predicted
for known satellite populations (see Mao \& Schneider~\cite{Mao98},
Chiba~\cite{Chiba01}).
This can be consistent with CDM and the lower density of
Galactic satellites if star formation is suppressed in most such
satellites as already discussed by Klypin et al.~(\cite{Klypin99})
and Bullock et al.~(\cite{Bullock00}).
For the $dn/dM \propto M^{-1.8}$ ($M_{\rm low} < M < M_{\rm high}$)
mass function expected for satellites (e.g. Moore et al.~\cite{Moore99},
Klypin et al.~\cite{Klypin99}) our test provides a rough estimate of the
upper mass scale $M_{\rm high}\simeq 10^6$--$10^9 M_\odot$.  While
this is uncomfortably close to the masses capable of disrupting stellar
disks and globular clusters (e.g. Moore et al.~\cite{Moore99}),
Font et al.~(\cite{Font01}) find that the expected CDM
substructure is consistent with the survival of thin galactic disks.
Thus, our result confirms a surprising
if generic prediction of CDM models and can be regarded as a major
success of the CDM model.  By the same token, alternatives to
CDM which aim to suppress small-scale power (warm dark matter)
or to destroy small satellites (self-interacting dark matter)
are accordingly disfavored. 

We believe that three other explanations, systematic errors in the data,
unmodeled, coherent structures in the lens and stellar microlensing, 
are unlikely.  While there are systematic errors in the
lens data, the anomalous flux ratios which drive the detection of 
substructure are present at levels far above the measurement errors and 
appear in multiple observations at differing wavelengths over periods of 
years.  They may be misinterpreted but not eliminated.  They are
also unlikely to be due to coherent structures in the lens galaxy.  
While we analyzed the lenses using singular isothermal ellipsoids
in an external shear for the macro model, Metcalf \& Zhao~(\cite{Metcalf02})
have shown that the flux ratios cannot be explained by a broad
range of macro models. 
The typical lens galaxy, including all seven discussed here,
is an early-type galaxy whose surface brightness profile is
well modeled by a smooth, elliptical de Vaucouleurs profile
(e.g. Lehar et al.~\cite{Lehar00}, Kochanek et al.~\cite{Kochanek00})
with no obvious photometric residuals.  Coherent features in the
lenses like spiral arms would be trivially detected in most
cases.  Moreover, if we need $f_{sat} \sim 0.01$ in compact components
like satellites to perturb the images, we would require a far bigger 
mass fraction in large scale, coherent structures that cannot produce 
perturbations isolated to a single image.  

The most problematic alternative explanation is stellar microlensing, which 
is the same physical phenomenon but produced by the stellar populations we
know to be present in the lens galaxy.  The basic argument against
microlensing is that it has too small a characteristic angular scale
($\mu$as=microarcseconds) to produce large, long-lived flux ratio anomalies 
given the sizes of typical radio sources.  The Compton limit, and direct VLBI 
observations the lenses, mean that typical sources are resolved on scales of
10--1000$\mu$as that are large enough to suppress the effects of
stellar microlensing.  The one apparent case of microlensing a
radio source, B1600+434, is probably due to a superluminal 
sub-component of the radio source where Doppler boosting gives
the source a smaller effective size and a rapid modulation
time scale (see Koopmans \& de Bruyn~\cite{Koopmans00}). Even in 
B1600+434, microlensing provides only a $\sim$5\% rms variation in the 
fluxes.  Moreover, many of the radio lenses also have constant flux
ratios on long time scales (years) which are difficult to reconcile 
with producing flux ratio anomalies using the stars.  Finally, our
method provides an estimate for the characteristic mass scale which
is grossly inconsistent with stellar microlensing.  This is
reinforced by detailed analyses of B1422+231 (Keeton~\cite{Keeton02},
Bradac et al.~\cite{Bradac02}) which find mass scales
compatible with CDM substructure but not stellar microlensing.
In summary, satellites are the most
natural explanation, and the required densities are comparable to
that expected in CDM and higher than that observed in normal
satellite populations.  The lenses cannot address directly whether 
they are dark or luminous because of the enormous distances.

Our examination of the problem is a preliminary one, and our estimates
can be extended and improved if the following points are addressed. 
First,  the entire question of the image fluxes and their uncertainties needs
to be carefully reconsidered.  We used a fixed measurement error of
10\% for the image fluxes, but the estimated surface density and its
uncertainties are affected by differences between the true errors
and the errors used in the analysis.  Until now there has been 
little motivation for determining image flux ratios with high 
precision (say 1\% accuracy), but improved analyses will need such
high precision.  Lens monitoring and time delay measurements, 
already important for using the lenses to determine the Hubble
constant without the systematic problems of the local distance
scale (e.g. Schechter~\cite{Schechter99}), are needed to 
eliminate the effects of source variability on the flux ratios.  
In optical lenses,
observations over a broad range of wavelengths are needed to provide
accurate corrections for extinction (see Falco et al.~\cite{Falco99}).  

Second, improved observations of the lenses are needed.  The lensed
images of the host galaxies of the radio sources, which are relatively
easy to observe using deep infrared imaging with HST, can be used
to constrain the macro model (e.g. Kochanek, Keeton \& McLeod~\cite{Kochanek01a}). 
Unlike the unresolved images of quasars or the marginally resolved images
of radio cores, large lensed structures like the host galaxies 
($\Delta\theta \gtorder 0\farcs1$) constrain the macro model without
being affected by substructure.  Combining the large scale constraints
with the compact images allows us to probe the substructure while
limiting the ability of the macro model to mask its effects.
Simultaneously, very high resolution, high dynamic range VLBI observations 
to map thin, extended radio structures can be used to extend the
search for substructure over larger regions in each lens 
(e.g. Wambsganss \& Paczynski~\cite{Wamb92}, Metcalf \& Madau~\cite{Metcalf01}).  
If the VLBI observations can show that the anomalous flux ratios are 
consistent with the geometric structure of the image, then we can
completely rule out microlensing as an alternative explanation.  Finally,
careful searches for additional, but faint, VLBI images produced by
the substructure may be a powerful means to constrain the density 
profiles of the satellites.  We have already found that our assumed
density satellite profile occasionally produces additional, detectable
images, suggesting that a shallower density profile would be preferred.

Third, the analysis can be expanded to include complete treatments of
the mass spectrum, the density profiles of the substructure and the
effects of finite sized sources.  These additional complications were
unwarranted in this first calculation because with only 7 lenses all
we can realistically say we have measured are the average properties
of the substructure. Any model producing the same average shear and
astrometry perturbations should be consistent with the data.  It
is clear from our Monte Carlo simulations, where our model would
occasionally generate additional images, that the density distribution
of the more massive substructures can be constrained by limits on the 
production of extra images.  Given our estimated angular scales for
the substructure perturbations and the dominance of the mass spectrum
by the higher mass halos, our effects should be little affected by
finite sources sizes.  If the typical radio source is 1~mas, then
we are modestly underestimating the surface density.

Finally, larger samples of lenses can reduce the considerable Poisson uncertainties. 
At least two additional radio quads have been discovered 
(B0128+437, Phillips et al.~\cite{Phillips00};
and B1555+375 Marlow et al.~\cite{Marlow99}) in the CLASS survey
we used as the basis for our analysis, but lack the HST
imaging data needed to accurately determine the position of the
lens galaxy.  Two-image lenses, while less optimal because of their
lower average magnifications, can be included in the analysis when
additional lensed structures like the images of the quasar host galaxy
or VLBI subcomponents provide the constraints needed to
break the degeneracies between the macro model and the substructure
we expect for a simple two-image lens.

Lastly, we note that other probes of substructure may be possible in
Local Group galaxies.  Very recently, Ibata et al.\
(\cite{Ibata01a,Ibata01b}) have suggested that the paucity of tidal
streamers in the Milky Way halo may betray the presence of
halo substructure.  Johnston et al. (\cite{Johnston01}) similarly
analyze tidal debris from the disrupted Sagittarius dwarf, and find
that stars in these tidal tails appear to be more scattered than
expected for debris orbiting in a smooth halo.  Thus there are
tantalizing hints of evidence for substructure within our own halo,
and further work along this avenue may lead to more definite
conclusions than is presently possible. Whatever the outcome
of these local studies, however, only gravitational lenses can
detect directly CDM satellites in which star formation was
suppressed.

\section*{Acknowledgments}
We thank David Rusin, Josh Winn and Stuart Wyithe for many helpful 
discussions.  We also thank Charles Keeton and David Rusin for making 
their lensing codes available, and Joanne Cohn for suggesting ND
visit SAO.  CSK was supported by the Smithsonian 
Institution and NASA grants NAG5-8831 and NAG5-9265.  ND was supported 
by the Smithsonian Institution Short Term Visitor Program, the
Dept. of Energy under grant DOE-FG03-97-ER 40546, and the ARCS Foundation.

\end{document}